# Be,La,U-rich spherules as microtektites of terrestrial laterites: What goes up must come down

Steve Desch[1]

[1]School of Earth and Space Exploration, Arizona State University, PO Box 876004, Tempe AZ 85287-6004

**Abstract**

Recently Loeb *et al.* (2024, "Recovery and Classification of Spherules from the Pacific Ocean Site of the CNEOS 2014 January 8 (IM1) Bolide", *Res. Notes Amer. Astron. Soc.* 8, 39) reported the magnetic collection of millimeter-sized spherules from the seafloor near Papua New Guinea. About 22% had Mg/Si < 1/3 and were identified as a new "differentiated" variety of cosmic spherule ("D-type"). In a subset of 26 of these "D-type" spherules, 12 "BeLaU" spherules were found to be dominated by Fe and Al, marked by low Si and even lower Mg content, depletions of volatiles species like Pb and Cs, and remarkable enrichments of Be, La, U, Ba, and other elements. Loeb *et al.* claimed these have exotic compositions different from other Solar System materials. We show that in fact samples with these compositions are not just found on Earth, they are *from* Earth; specifically, we identify them as microtektites of terrestrial lateritic sandstone. Based on the location of the sample site, we associate them with the Australasian tektite strewn field, generated 788 kyr ago by an impactor that melted and ejected ~$10^8$ tons of sandstone, including a lateritic layer, from Indochina. A tektite origin for the spherules is corroborated by their terrestrial Fe isotopic compositions and the compound, non-spherical nature of many of them, which preclude formation as ablation spherules from a bolide. Due to the restriction of laterites to the tropics, iron-rich tektites may be uncommon, but we predict they should comprise ~3% of the Australasian microtektites.

## INTRODUCTION

In August 2023, Prof. A. Loeb led a costly oceanographic expedition to collect about 850 millimeter-sized spherules from the seafloor near Papua New Guinea, under the presumption that many would derive from the CNEOS 2014-01-08 bolide, which they asserted was interstellar in origin. As reported in unrefereed work[1,2], 745 that are not obviously terrestrial were chemically analyzed by X-ray fluorescence (XRF). About 78% of them were found to resemble typical micrometeorites ablated from iron or chondritic meteoroids. The remaining 162, or 22%, have molar ratios Mg/Si < 1/3, and were called "D-type" by Loeb *et al.* (2024). Of these "D-type spherules", 26 were analyzed by inductively coupled plasma mass spectrometry (ICP-MS) for certain trace elements, and 12 were found to have enrichments in Be, La, U, and other elements. Loeb *et al.* named these "BeLaU" spherules. Besides these enrichments, they are iron-rich (19-55wt% Fe) and aluminum-rich (10-45wt% $Al_2O_3$), and they are depleted relative to chondritic (CI) or Earth (upper continental crust) compositions in the volatile elements Cs and Pb. Despite appearing melted, they were found to lie outside the range of common terrestrial igneous rocks in a Mg-Al-Fe ternary diagram. Based on the depletion of the highly siderophile element Re, Loeb *et al.* (2024) concluded they derived from a differentiated planet, but also stated the "BeLaU" spherules "appear to have an exotic composition different from other Solar System materials."

---

[1] Loeb et al. (2023): https://arxiv.org/pdf/2308.15623.pdf
[2] Loeb et al. (2004): https://iopscience.iop.org/article/10.3847/2515-5172/ad2370

In order for the spherules to be interstellar in origin, they would have to be ablation spherules from the 2014-01-08 bolide, *and* the 2014-01-08 bolide would have to be interstellar in origin. Collection of extrasolar material would be one of the most significant discoveries in meteoritics and planetary science, but an interstellar origin for the bolide has been widely disputed. Comparison of dozens of meteor trajectories observed with high precision from ground camera networks, with bolide velocities reported by the Department of Defense to the Jet Propulsion Laboratory Center for Near-Earth Object Studies (JPL-CNEOS), show that the reported velocities entail significant (up to ~8 km/s) uncertainties (Brown *et al.* 2016; Devillepoix *et al.* 2019). It has been concluded that the 2014-01-08 bolide was in a bound, heliocentric orbit and simply appeared to have been on an interstellar trajectory because of mismeasurement (Brown and Borovicka 2023; Hajdukova *et al.,* submitted to *A&A*). The identification as interstellar is essentially a 3σ result, and 3σ results are wrong one in a thousand times; the CNEOS catalog, notably, contains about one thousand meteors. The "BeLaU" spherule S21 also is compound, comprising three attached spherules. Ablation spherules are always singular objects, formed from completely liquid melt droplets; they are not compound objects (van Ginneken *et al.* 2021), nor are compound spherules predicted to form from ablated material. The S21 spherule—and by extension other "BeLaU" spherules—must have a distinct origin not associated with the 2014-01-08 bolide. These issues, and many other fatal flaws of the interstellar hypothesis, were reviewed by Desch and Jackson (2023)[3].

**Iron Isotopes Indicate an Origin on Earth**

In the earlier, unrefereed work uploaded to arXiv, Loeb *et al.* (2023) also reported measurements of the Fe isotopic ratios of two "BeLaU" spherules, "S21" and "S10". In a plot of $\delta^{57/54}$Fe vs. $\delta^{56/54}$Fe, both were found to lie exactly (to within < 0.1‰ uncertainty) on the terrestrial fractionation line (TFL) on which all materials from the Solar System plot, including those from Earth. The spherule S10 lies at $\delta^{56/54}$Fe = -0.2‰, and S21 at $\delta^{56/54}$Fe = +1.1‰. Loeb *et al.* (2023) claimed that their departure from $\delta^{57/54}$Fe ≈ $\delta^{56/54}$Fe ≈ 0‰ was indicative of Rayleigh distillation of Fe during entry, although it must be noted that ablation spheres as iron-rich as S10 and S21 generally show much larger fractionations ($\delta^{56/54}$Fe ≈ +20‰ to +36‰) (Engrand et al., 2005), and not $\delta^{56/54}$Fe < 0‰ like S10. As reviewed by Desch and Jackson (2023), the case for isotopic fractionation during atmospheric entry is ambiguous, but the fact that these spherules are so close to the terrestrial isotopic ratios means that an interstellar origin can be effectively ruled out.

If the 2014-01-08 bolide hypothetically originated in another solar system, it would contain iron reflecting the time and place of formation in the Galaxy of that system. Even if it were from the same galactocentric radius as the Sun, the other system could have formed at any random time over the age of the Galaxy. Over the last 7 Gyr, the average metallicity of stars has increased from $Z \approx 0.5\ Z_{SUN}$ to $Z \approx 1.3\ Z_{SUN}$ (Carrillo *et al.* 2023). According to Galactic Chemical Evolution models, such stars should form with a range of isotopic ratios $\delta^{56/54}Fe \approx$ -100‰ to +1500‰, and $\delta^{57/54}Fe \approx 0.9\ \delta^{56/54}Fe$ from -100‰ to +1300‰ (Kobayashi *et al.* 2011). This range is depicted in **Figure 1**. The probability of material in another solar system having $\delta^{56/54}Fe$ matching terrestrial to within 1‰ would be ~$10^{-3}$, and the probability of it further having $\delta^{57/54}Fe$ matching terrestrial to within 0.1‰ would be ~$10^{-1}$. We conclude that > 99.99% of all solar systems from which an interstellar meteor would derive have measurably distinct Fe isotopic ratios. There is a < 0.01% chance the S10 and S21 spherules—and by extension all "BeLaU" spherules—are interstellar, if the Fe isotopic measurements reported by Loeb *et al.* (2023) were verified.

---
[3] https://arxiv.org/ftp/arxiv/papers/2311/2311.07699.pdf

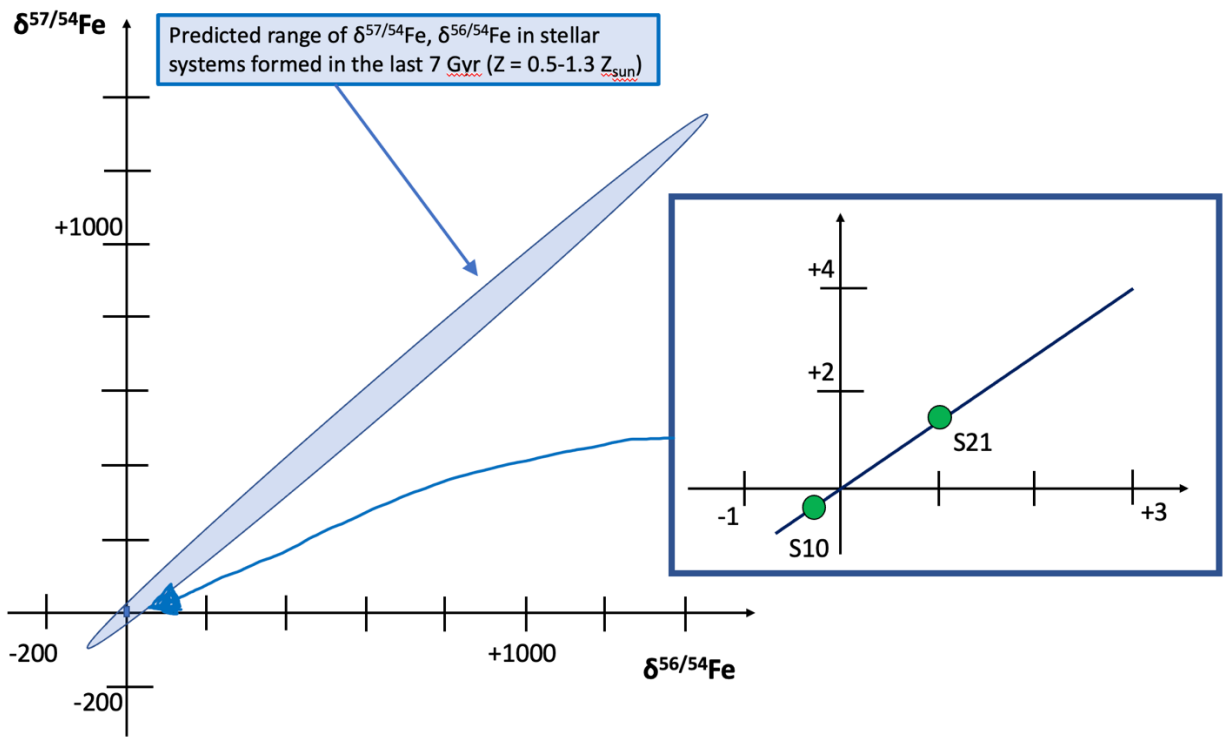

**Figure 1.** Range of iron isotopic ratios of the spherules collected by Loeb et al. (2023; arXiv 2308.15623), compared to the predicted range among stars born in the last 7 Gyr of the Galaxy, the sources of hypothetical interstellar meteors. There is a < $10^{-4}$ probability that iron from another solar system would match the Earth's composition so exactly.

**Compositions Indicate an Origin on Earth**

There is every indication that the "BeLaU" spherules are terrestrial in origin: the Fe isotopes of S10 and S21 lying on the TFL diagnose them as from the Solar System; and the compound nature of S21 and its small magnitude of Fe isotopic fractionation both argue against it being an ablation spherule from an incoming meteoroid. It therefore must be from Earth itself. However,

Loeb *et al.* (2024) have continued to speculate about an interstellar origin for the "BeLaU" spherules because of their compositions. They assert the "BeLaU" spherules are depleted in Re, and therefore arise from some differentiated planet with a core, and that their low Mg/Si ratios especially "reflect a highly differentiated, extremely evolved composition, of an unknown origin." They are describing Earth—the planet on which the spherules were collected—yet conclude that the spherules "appear to have an exotic composition different from other Solar System materials." It seems a specific terrestrial source and a history for the spherules are demanded.

Strong hints for their origin come from the data already presented by Loeb *et al.* (2024). They used ICP-MS to measure a suite of 56 elements in 12 "BeLaU" spherules and scaled the concentration of each element in the spherules to the concentration of that element in CI chondrites, a standard reference thought to represent the starting composition of the Solar System. Whereas the concentration of Fe in the spherules is only a few (1-3) times the concentration of Fe in CI chondrites, the concentration of U is 300-2000× the U concentration in CI chondrites, with similar enrichments for Be and La. Loeb *et al.* (2024) compared the spherules to other chondritic or igneous samples but didn't find any with such high concentrations. Igneous meteorites from Mars or Vesta, for example, are enhanced in U by factors of only ≈20. The KREEP component from mantle-crust differentiation on the Moon is enhanced by 700× relative to CI, but is not a good match to other elements, especially Mg, which is low (0.05× CI) in the spherules but present at levels 0.5× CI in KREEP.

However, Loeb *et al.* (2024) also compared S21 to the average composition of Earth's Upper Continental Crust (UCC). We have created a similar plot, displayed in **Figure 2**, using the UCC compositions of Rudnick and Gao (2014) and using the arithmetic means of the concentrations reported by Loeb *et al.* (2024) for each "BeLaU" spherule. We note some differences between this plot and Figure 20 of Loeb et al. (2024): relative to UCC, we find that Zn is enriched, not depleted, and that Sb and W are more enriched than they found. We cannot explain the differences. Overall, though, the trends are similar, and undoubtedly UCC is a much better match than CI to the average of the "BeLaU" spherules.

In a similar manner, Gallardo (2023) demonstrated that anthropogenic coal ash is a good fit to the average composition of the spherules, although they are somewhat depleted in volatiles like As, Se, Rb, Cd, Sb, Cs, Tl, Pb, and Bi, a point made by Loeb *et al.* (2024). Still, coal ash, like UCC (because it largely reflects UCC in its trace element composition) is also a much better match than CI to the average of the "BeLaU" spherules, reinforcing that the spherules appear to be of terrestrial origin.

It is apparent that relative to UCC, Mg, Si, and the alkalis Na, and K are quite depleted. For illustrative purposes, we arbitrarily assume 2/3 depletion of major elements, so that trace elements are enriched by a factor of 3. We then find that the concentrations of almost all elements in the spherules match those in UCC, to within a factor of 2. The only exceptions are a few that are more enriched (Be, Fe, Zn, Mo, Sb, and W) and certain volatile elements that are more depleted (Rb, Cs, Tl, Pb, Bi, and As). While UCC is not an exact match to the "BeLaU" spherules, either, it is a significantly better match than CI. This is a strong hint that the source materials of the "BeLaU" spherules are terrestrial. Specifically, they likely derive from *weathered* crustal materials from which MgO, SiO$_2$, Na$_2$O and K$_2$O have been leached, which were later heated so that elements like Rb, Cs, and Pb devolatilized.

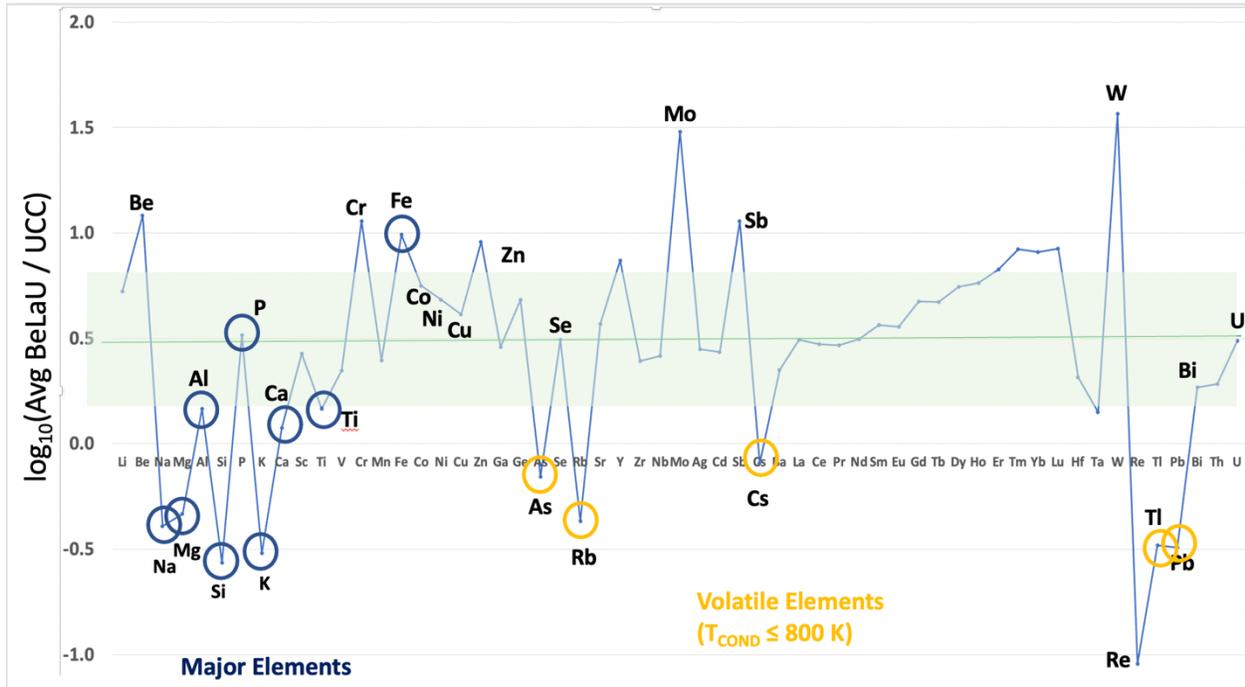

**Figure 2**. Average elemental concentrations in "BeLaU" spherules, reported by Loeb et al. (2024), scaled to the abundances in Earth's upper continental crust (UCC; Rudnick and Gao, 2014). We highlight in blue the major elements, showing strong relative depletions in Mg, Si, Na, and K; the spherules are dominated by iron and aluminum (or their oxides). We also highlight (in yellow) certain volatile elements, most with condensation temperatures below 800 K (Lodders 2003). Trace elements within the spherules are broadly consistent (to within about a factor of 2, green band) with loss of ~2/3 of the mass of UCC (nearly complete loss of SiO$_2$, MgO, Na$_2$O and K$_2$O) and passive enrichment of other species by a factor of 3, green line, followed by strong loss of some volatiles. While not an exact match, this strongly hints at a source material that is from Earth's surface, but weathered, and then partially melted.

**The Spherules are Moderately Melted Earth Rock**

As highlighted by Figure 2, most elements in the "BeLaU" spherules appear (passively) enriched over UCC by factors ~3, excluding some elements with low condensation temperatures $T_{COND}$ < 800 K (as tabulated by Lodders, 2003). It should be remembered that the $T_{COND}$ are the temperatures below which > 50% of the element exists in solid form, in a gas of solar (or CI chondrite) composition. The volatilization of many species is sensitive to oxygen fugacity, which is very different in a melted crustal rock near the iron-wüstite buffer vs. a solar-composition gas some six orders of magnitude less oxidizing. Nevertheless, the only trace elements depleted significantly (relative to a factor-of-3 enrichment) are As, Rb, Cs, Pb, and Tl, most of which have $T_{COND}$ ≤ 800 K. (Re is of very low abundance, and difficult to measure.) It is reasonable to conclude that the spherules, on average, spent time near or above 800 K, and devolatilized. However, the non-depletion of similar volatile species like Se (697 K), Ga (883 K), and Ge (968 K) suggests temperatures were not far above 800-1000 K in these spherules for very long. For comparison, iron melts at 1265 K and silica at 1437 K, suggesting the spherules were only partially molten.

Other morphological evidence suggests the spherules were partially, but not completely melted. Figures 3 and 4 of Loeb *et al.* (2024) show backscatter electron images of various spherules. The typical I-type, S-type and G-type micrometeorites formed by ablation of meteoroids are clearly perfectly round, the shape expected for completely melted liquid droplets. In contrast, the "D-type" spherules are very frequently compound, including spherule S21, which appears to be three spheres stuck together. Had this been fully molten, it would have coalesced into a single round sphere. Most of the "D-type" spherules are distinctly *non-spherical*. Spherule 17MAG-2 (a low-Si "BeLaU" spherule) is in fact cuplike in morphology.

**The Spherules as Microtektites of Lateritic Sandstone**

The above lines of evidence—all of which were presented by Loeb *et al.* (2024)—should lead one to the conclusion that the "BeLaU" spherules are partially melted bits of Earth rock, i.e., that they are *microtektites*, millimeter-sized bits of Earth rock melted and flung thousands of kilometers away during impacts. It can be further surmised that the target rock was broadly consistent with Earth's crust, yet depleted in Mg, Si, Na, and K, and enriched in iron and aluminum. Given the geology of the location where the spherules were collected, *laterites* are a natural candidate for the target rock.

Laterites are heavily weathered soils, found mostly between 20°S and 20°N latitude, whose surface layers have been markedly depleted in alkalis (e.g., $Na_2O$), $SiO_2$, and especially MgO (Jha et al. 2019), among other species, by the intense rainfall in the tropics. These layers are often so dominated by iron-rich (hematite and goethite) and/or aluminum-rich (gibbsite, boehmite, diaspore, and kaolinite) minerals that they are mined, e.g., the world's largest bauxite mine, Weipa, in northern Australia, only 1400 km from where the spherules were collected. Laterites even exist on nearby Manus Island (Owen, 1954). The topmost meters of lateritic sandstones can exhibit remarkable (factors of 3× to 30×) enrichments in Cr, V, Ni, Ba, Sc, Be and U, and large enrichments of light rare earth elements (LREEs) like La over heavy rare earth elements (HREEs) (Jha *et al.* 2019), relative to their host rocks, naturally explaining the peculiar compositions of the spherules. **We hypothesize that the "BeLaU" spherules are microtektites of the uppermost, iron- and aluminum-rich layers of lateritic sandstones.**

An obvious and severe test of our hypothesis is whether the site near Papua New Guinea (1.3°S, 147.6°E) where Loeb *et al.* collected their spherules is near the site of a tektite-producing impact that hit target rock of lateritic sandstone. As it happens, one of the largest, most recent, and best-studied tektite fields in the world is the Australasian tektite strewn field created by an impact in Indochina, 788 kyr ago (Jourdan *et al.* 2019). Analysis of the Australasian tektites overwhelmingly shows the target rock was sandstone with specific properties. The concentration of tektites indicates the impact was in Indochina. The target rocks there are sandstones consistent with the tektites, and the tektites there are found to have been deposited on a horizon of lateritic sandstone (Tada *et al.* 2020). This definitively demonstrates that microtektites of lateritic sandstone should have been produced by the Australasian impact. Microtektites from this event were deposited up to 12,000 km away, reaching Madagascar, Antarctica, Australia, and the Philippines. The strewn field includes the spherules collection site, 5000 km away.  This is an overwhelming confirmation of the viability of the hypothesis that the "BeLaU" spherules are microtektites of lateritic sandstone.

**Outline**

The rest of this paper is devoted to further, more-detailed tests of the hypothesis. In the section *How Many Microtektites and Microtektites of Lateritic Iron?*, we discuss the Australasian tektite field in the context of our hypothesis. We are specifically interested in constraining the number of microtektites from lateritic sandstone that would have been deposited and magnetically retrieved from the collection site. Could "D-type" spherules comprise 22% of the spherules collected there? Could a significant fraction of these be "BeLaU" spherules?  In the section *Geochemistry of "BeLaU" Spherules vs. Microtektites of Lateritic Iron,* we test whether the chemical abundances of target rocks in the impact site, or other tektites, provide a compelling match to the compositions of the spherules. We discuss how laterites match the major-element compositions of the spherules, but we especially focus on the relative abundance patterns of REEs and trace elements. We summarize our findings in the *Conclusions* section. We predict that up to ≈3% of Australasian microtektites may be iron-rich, possibly mistaken for I-type cosmic spherules. We predict that an effort to magnetically collect of seafloor spherules, similar in methodology and scale to the effort made by Loeb *et al.,* but in control areas hundreds of km to the WNW, should find similar numbers of "BeLaU" spherules. These spherules, like other the "BeLaU" spherules, obviously would have nothing to do with the 2014-01-08 bolide, and would in fact be from Earth.

## HOW MANY MICROTEKTITES AND MICROTEKTITES OF LATERITIC IRON?

**The Australasian Tektite Strewn Field**

To contextualize our hypothesis and estimate the number of "BeLaU" microtektites at the Papua New Guinea collection site, we review the Australasian tektite strewn field. Tektites are derived from terrestrial rock that is melted and flung outward by a meteorite impact, resolidifying before hitting the ground. Tektites typically range in size from submillimeter (microtektites) up to centimeters in size or larger. The compositions of tektites reflect the composition of the target rock, possibly mixed with some meteoritic impactor material. Because the target rock is, in general, heterogeneous, the compositions of tektites and microtektites can be quite variable, even if they originate in the same impact.

All impacts can generate tektites, and many examples are known, but there are just a few major tektite strewn fields on Earth: one in North America, associated with the 85 km crater in Chesapeake Bay from 35.5 Myr ago (Koeberl *et al.* 1996); another strewn field of predominantly clinopyroxene spherules encompassing most the globe, apparently associated with the 100 km Popigai impact structure in northern Siberia from 35.5 Myr ago (Bottomley *et al.* 1997) [the impactors, both at the end of the Eocene and the start of the Oligocene, may have been contemporaneous]; one associated with the 24 km Ries impact structure in central Europe from 14.8 Myr ago (Schwarz *et al.* 2020); one near Ivory Coast, Africa, associated with the 10.5 km Bosumtwi impact crater from 1.07 Myr ago (Cuttitta *et al.* 1972); and the Australasian strewn field.

The Australasian strewn field is for practical purposes the largest tektite strewn field in the world (after that from Popigai), and the most recent. Tektites from this impact are Ar-Ar dated to 788.1 ±3.0 kyr ago (Jourdan *et al.* 2019). **Figure 3** taken from Folco *et al.* (2023), illustrates the distribution of the Australasian tektites and microtektites ("AATs"). Their range extends across up to 30% of the Earth's surface: south 11,000 km to Antarctica (van Ginneken *et al.* 2018; Soens *et al.* 2021), across the Australian continent, west across the Indian Ocean to near Madagascar, and east into the Pacific, possibly as far as 10,000 km away to western Canada (Schwarz *et al.* 2016). The range, notably, includes the location 5000 km away at 1.3°S, 147.6°E, about 85 km north of Manus Island, Papua New Guinea, where Loeb *et al.* collected their spherules.

Unlike other tektite strewn fields, no associated crater has been definitively located, although much is known about it. Based on the inferred total mass of AATs (~$10^8$ tons; Glass 1990), a recent estimate of the impactor diameter is 1-2 km, producing a crater 43 ± 9 km in diameter (Glass and Koeberl, 2006); presumably the crater of the large Australasian tektite strewn field would be comparable in size to the craters associated with the other tektite strewn fields above. The crater must be in southeast Asia. The densities of tektites and microtektites increase with proximity to southeast Asia (Glass and Pizzutto, 1994; Glass and Koeberl, 2006; Prasad *et al.* 2007). The largest (tens of cm) AATs, so-called Muong Nong layered tektites, are found exclusively in Cambodia, Vietnam, southern Laos, northeastern Thailand, and Hainan Island, China (Tada *et al.* 2020). These were possibly ejected in specific directions, or rays, from an impact site in the Gulf of Tonkin (Whymark 2013). The concentration of $^{10}$Be in AATs (a measure of the depth from which the tektites were launched) increases systematically with distance from Indochina, consistent with shallower depths being launched farther. The lowest $^{10}$Be abundances are found in a zone spanning southern Laos, the Gulf of Tonkin, and Hainan Island (Ma *et al.* 2004).

The target rock was undoubtedly sediments. Muong Nong tektites reached lower temperatures than other AATs and often contain unmelted relict grains of quartz and zircon, indicating the source material for the AATs was fine-grained, silt-sized (4-63 µm) clastic sediments similar to shale or greywacke, with contributions from alkali-bearing feldspar and plagioclase, Ti-bearing phases (rutile, ilmenite and titanite), magnetite, and carbonates like calcite and dolomite (Glass and Barlow, 1979; Glass and Koeberl 2006; Glass and Fries 2008). Pb isotopes indicate three sources (perhaps feldspar, zircon, and organic matter adsorbed on sediments), sorted during fluvial transport (Ackerman *et al.* 2020). Nd and Sr isotopic systematics of AATs reveal the sediments were derived from Mesozoic to Lower-Paleozoic succession of sedimentary and magmatic rocks that were then redeposited in the late Neogene and/or early Quaternary period into another sedimentary basin that was the target (Blum 1992; Ackerman *et al.* 2020). Based also on variations in the $K_2O$/CaO ratios and their correlation with $^{87}Sr/^{86}Sr$, Ackerman *et al. (*2020) and Soens *et al.* (2021) concluded the target rock was chemically and mineralogically stratified and rich in plagioclase and/or carbonates. The relatively low K/Na ratio of AATs argues against an impact in a heavily vegetated area (Mizera *et al.* 2016). AATs record high and variable abundances of $^{10}Be$, indicating they derived from the uppermost tens of cm of sedimentary rock exposed to meteoric $^{10}Be$ (Ma *et al.* 2004).

Based on the above evidence, an impact in the continental shelf in present-day Gulf of Tonkin, perhaps the Song Hong-Yinggehai basin, appears favored (Ma *et al.* 2004; Rochette *et al.* 2018; Whymark 2013, 2018; Ackerman *et al.* 2019, 2020). While underwater today (and for most of the last million years), at the time of the impact sea levels were transiently about 100 m lower than today (Pillans *et al.* 1998; Kominz 2001), meaning that most of what is now the Gulf of Tonkin would have been exposed and subaerial. This potentially could explain why the impact site was relatively unvegetated land, heavily and uniformly sedimented yet not mostly marine. It also potentially explains the non-discovery of the crater, as it would be buried today beneath hundreds of meters of sediment (Luo *et al.* 2003; Whymark 2016). The sediment flux from the Red River alone (tens of Mtons/yr; Quang and Viet 2023) is sufficient to add hundreds of meters of sediment to the floor of the entire Gulf of Tonkin, sufficient to bury evidence of even a crater 85 km in size, with rims hundreds of meters high. Although it has recently been suggested that a crater exists buried under 300 m of basalt on the Bolaven Plateau in southern Laos (Sieh *et al.* 2020), the small size of the putative crater (15 km), as well as other geochemical evidence, appear inconsistent with this locale (Mizera *et al.* 2022).

Numerical simulations show that tektites are launched efficiently by high-velocity (>> 15 km/s) and oblique (30-50° from the horizontal) impacts into silica-rich target rocks, as a "melt sheet" of the topmost ~$10^2$ m is shocked outward (Artemieva 2001; Stöffler *et al.* 2002; Goderis *et al.* 2017). The large "layered" Muong Nong tektites may have formed by differential flow of melts with different viscosities in a melt sheet after an inhomogeneous melt reached the surface (e.g., Barnes and Pitakpaivan 1962; Barnes 1963; Koeberl 1992; Wasson 2003; Goderis *et al.* 2017). This model is supported by paleomagnetic measurements of the alignment of magnetizations in Muong Nong tektites (Gattaccecca *et al.* 2022). A prediction of the above numerical models of the AATs is that the deepest strata were not generally launched more than a ~$10^3$ km, and the farthest-launched material, reaching 11,000 km away, came from very shallow layers. This is corroborated by the decrease in mean particle size with distance from Indochina, with microtektites in Antarctica limited to < 500 μm in size (van Ginneken *et al.* 2018; Soens *et al.* 2021), compared to centimeter-sized australites in Australia and Muong Nong tektites tens of cm in size, consistent with simulations. Different depths of launching also are suggested by the trends of increasing Sr and decreasing $Na_2O$ and $K_2O$ content with distance from Indochina (Soens *et al*. 2021), and by the generally increasing concentrations of

$^{10}$Be in AATs with increasing distance from southeast Asia. These vary from 59 ×10$^6$ atoms/g for a layered tektites found in Thailand, to 280 ×10$^6$ atoms/g in splash-form tektites in Australia (Ma *et al.* 2004), or 184 × 10$^6$ atoms/g for microtektites in Antarctica (Rochette *et al.* 2018), after correcting for *in situ* production and dating back to the time of the impact. The magnitude of $^{10}$Be concentrations is within the range of marine continental shelf deposits and lateritic bauxite in Taiwan (Ma *et al.* 2004). The large variations in $^{10}$Be concentrations practically demand excavation of different depths of materials, possibly over tens to hundreds of meters; Ma *et al.* (2004) estimated 15 to 300 m. Whatever was in the nearest tens of meters to the surface would have been launched by the impact.

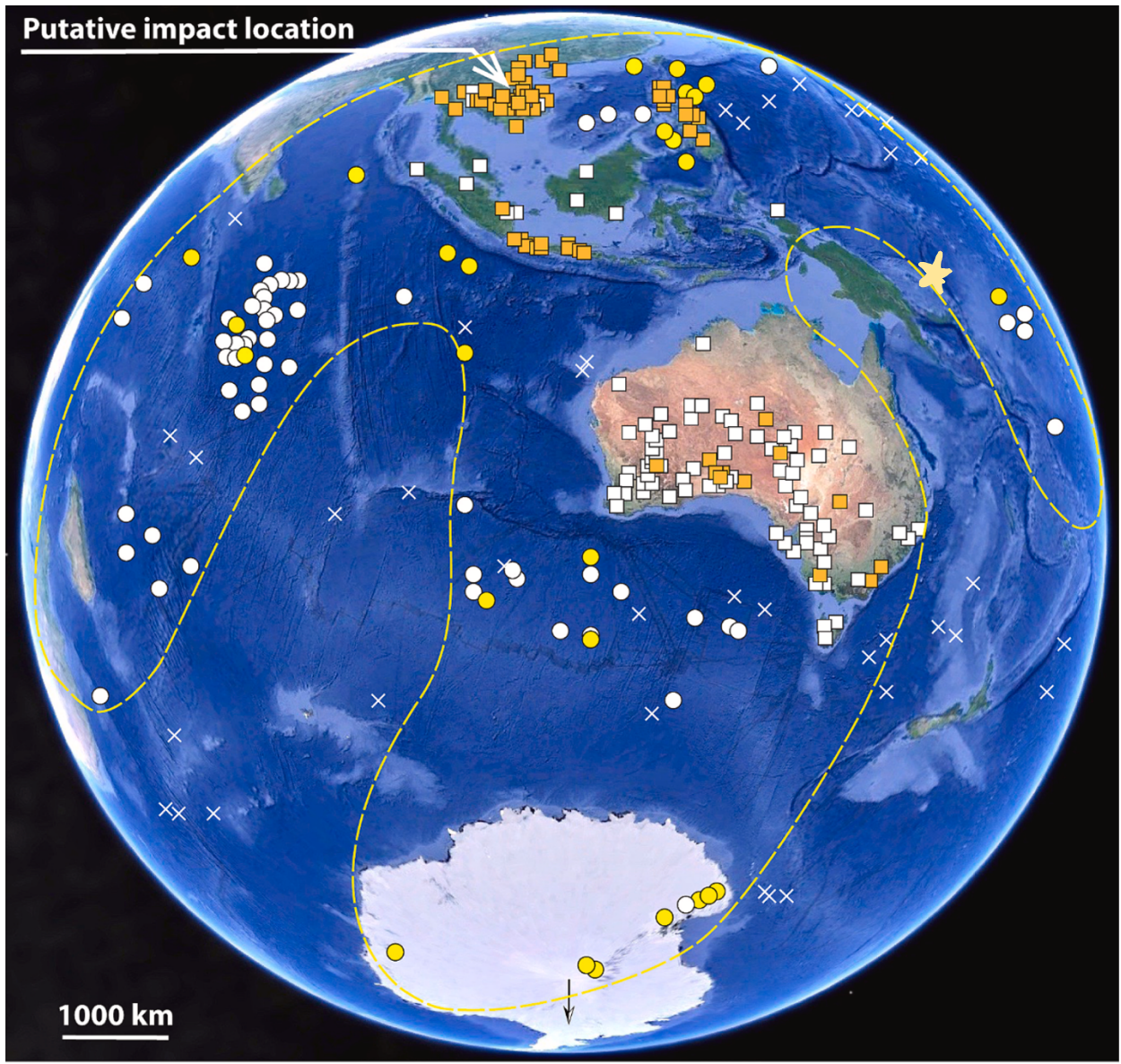

**Figure 3**. Approximate extent of the Australasian tektite strewn field. The yellow line demarcates regions where tektites from the presumed impact site (in the present-day Gulf of Tonkin) have been found (filled circles and squares) from regions where they have not been found (crosses). A yellow star marks the location where Loeb *et al.* collected spherules. Adapted from Folco *et al.* (2023).

**Microtektites of Laterites?**

Throughout Indochina, lateritic soils are common near the surface. Lateritic soils arise from weathering due to intense rainfall, which leaches some elements out of some strata and redeposits them in lower strata. Laterites can see concentrations to ore grade in some elements, especially alumina, and metals like Fe, Co, Ni, and Cu, over length scales of meters to tens of meters. Meters-thick layers of ferruginous concretions ('ferricrete') can overlie meters-thick layers of aluminous concretions ('bauxite'). Lateritic soils harden after being dug up and dried, and are commonly used in the tropics as bricks (*later* is Latin for brick), most notably in the Angkor Wat monument (e.g., Uchida 1999), demonstrating how common these soils are.

AATs have long been associated with laterites. The Muong Nong layered tektites are commonly contaminated by lateritic material (Fiske *et al.* 1996, 1999), and tektites are commonly found directly on top of a widespread lateritic layer across southeast Asia, now typically buried by 0.3 – 1 m of soil (Tada *et al.* 2020). Finding this lateritic layer is usually sufficient to locate AATs, as charmingly recounted by Nininger (1961). The study by Tada *et al.* (2020) of size distributions of tektites and their fragments makes clear that the AATs fragmented after falling onto the lateritic layer, which was then on the surface. These lines of evidence make clear that the impact was onto a lateritic sandstone, presumably with Fe-rich and Al-rich layers within meters of the surface, likely comprising the surface itself.

Australasian tektites and especially microtektites (because they sample smaller volumes) therefore should sample melts of lateritic Fe-rich or Al-rich layers. There is definitely metal in some AATs, but it is difficult to disentangle it from the metal from the impactor. Many indochinites and philippinites contain FeNi spherules that may have formed by *in situ* reduction of terrestrial iron, but which may derive from meteoritic metal (Chao 1964; Ganapathy and Larimer 1983). It is inferred from Ni-Mg mixing calculations and Co/Ni and Cr/Ni ratios that AATs contain metal that can be a mix (4-6wt% on average) of chondritic metal from an ordinary chondrite (or possibly an acapulcoite) and a terrestrial component broadly consistent with UCC but not a terrestrial mafic component (Goderis *et al.* 2017; Ackerman *et al.* 2019; Folco *et al.* 2018, 2023). However, the most Ni-rich AATs—those presumed to preferentially sample the impactor—are found concentrated to the south, on the islands of Borneo, Java, and Belitung (Goderis *et al.* 2017). No iron-dominated microtektites have been identified, but these would not be distinguishable from I-type cosmic spherules without chemical analysis. In contrast, Al-rich microtektites are common among AATs, and some have $Al_2O_3$ > 30-35wt% (Folco *et al.* 2016, 2023; van Ginneken *et al.* 2018). Although not identified as melts of bauxite, it is hard to imagine any way to obtain such a high alumina abundance in any tektite, even by volatile loss.

AATs were mostly melted and undoubtedly experienced devolatilization, but we do not attribute the high abundance of Al to passive enrichment due to volatile loss. Compared to

UCC, many microtektites (e.g., those from Antarctica) are highly depleted in the volatile elements Pb (727 K), Cs (799 K), and to some extent Zn (726 K). [Here again the temperatures are the 50% condensation temperatures of the element in a solar-composition gas, as calculated by Lodders (2003). These are sensitive to oxygen fugacity, so these temperatures provide only a rough guide to volatility.] Evaporative loss of Sn (704 K) (Creech and Moynier 2019), Pb (727 K) (Ackerman *et al.* 2020), and Zn (726 K) and Cu (1037 K) (Moynier *et al*. 2009, 2010; Rodovska *et al*. 2017; Jiang *et al.* 2019; Wimpenny *et al.* 2019) are also implied by their isotopic fractionation. In contrast, there is no evidence from isotopic fractionation for evaporative loss of Li (1142 K) (Magna *et al.* 2011; Rodovska *et al.* 2016) or K (1006 K) (Humayun and Koeberl, 2004; Jiang *et al.* 2019). It is not simple to disentangle vaporization losses of alkalis (Na and K) from variations in the target rock (see Soens *et al.* 2021), but given that $^{10}$Be and other elemental abundances correlate with distance from the impact due to stratification of the parent rock, and alkalis are stratified in weathered soils, we credit that for the observed variations of $Na_2O$ and $K_2O$, rather than devolatilization. We therefore conclude that devolatilization affected the abundances of elements with condensation temperatures up to only 800-1000 K, but variations in abundances of the major elements (Si, Mg, Fe, Ca, Ti), with much higher condensation temperatures > 1350 K, were not affected. We interpret the high (up to 35wt%) abundance of $Al_2O_3$ to be the ratio in the target rock, which *strongly* suggests that Australasian microtektites contain melts of lateritic soils. And, because aluminous laterites lie under ferruginous laterites, there should be about as many iron-rich microtektites as Al-rich microtektites.

**Numbers of Microtektites vs. Cosmic Spherules**

With this context, we estimate what fraction of spherules at the collection site at 1.3°S, 147.6°E should be large (> 1 mm) microtektites, and lateritic (Al- and/or Fe-rich) particles in particular. The collection site lies just under 5000 km from the presumed impact site in the Gulf of Tonkin. Other sites where Australasian microtektites have been found—near Madagascar, in Antarctica, and possibly Canada—are more than twice as far away from this site. The site is within where many authors (e.g., Soens *et al.* 2021; Pan *et al.* 2023; Folco *et al*. 2023) draw the strewn field boundaries. Compellingly, extension of the apparent impact rays over the Philippines (Whymark 2016) would intercept the exact location where the spherules were collected.

The total number density of Australasian microtektites > 125 μm in size recoverable (e.g., in drill cores) from the seafloor at this location is ≈20 $cm^{-2}$ (Glass and Pizzutto 1994; Lee and Wei 2000; Glass and Koeberl 2006). The size distribution of microtektites is such that the number > 1000 μm in size would be almost exactly 100 times smaller (Prasad and Sudhakar 1999), or 0.20 $cm^{-2}$. This is to be compared to the background number of cosmic spherules (micrometeorites) in deep sea sediments, which Prasad et al. (2013) estimate (after sieving 293 kg of sediment from the Indian Ocean) is 160 tons/yr, or 3 x $10^{-11}$ g $cm^{-2}$ $yr^{-1}$, globally. In the last $10^6$ yr this would correspond to 0.03 mg $cm^{-2}$, over 60% of which resides in particles > 300 μm in size. Assuming a median mass 0.04 mg suggests 0.7 $cm^{-2}$ background cosmic spherules. Within sediments ≈$10^6$ yr old, microtektites should make up a significant fraction—we estimate about 22%—of all spherules at the collection site.

Deriving from crustal rock, these 22% of spherules should generally have Mg/Si < 1/3 (UCC has Mg/Si ≈ 0.05; Rudnick and Gao 2014) and would be identified by Loeb *et al.* as "D-type". As it happens, out of the 745 spherules of seeming non-terrestrial origin analyzed by Loeb *et al.* using XRF, 162, or 22%, were identified as "D-type".

Loeb *et al.* further analyzed 26 of these "D-type" spherules using ICP-MS and identified 12 (46%) as "BeLaU" type. We associate these spherules with a specific Fe-rich lateritic layer that we estimate would comprise anywhere from 1-70% of microtektites, based on typical laterite thicknesses ≈3-10 meters, and an estimated thickness of the spall layer 15-300 m (Ma *et al.* 2004). Alternatively, the fraction of microtektites that are Al-rich is about 3% in multiple drill cores (1/37 ≈3%; Glass *et al.* 2004) and in two Antarctic locations (2/35 ≈3%; Folco *et al.* 2016; and 6/233 ≈3%; Folco *et al.* 2023). The fraction of Fe-rich "BeLaU" spherules would be a similar fraction ≈3%.

The frequency ≈3% of "BeLaU" spherules among D-type spherules we predict is much lower than the observed fraction 12/26 = 46% of D-type spherules analyzed by Loeb *et al.* using ICP-MS that were iron-rich "BeLaU" spherules, but this is not unexpected. Iron-rich spherules, because they can develop magnetite rims during atmospheric heating (Prasad *et al.* 2018), are much more easily collected magnetically. In deep-sea collections, I-type spherules comprise > 30% of all micrometeorites, despite comprising < 2% of all micrometeorites in unbiased Antarctic collections (Taylor *et al.* 2007). Most likely, iron-rich "BeLaU" spherules are overrepresented by a similar factor of 15 in the Papua New Guinea sample, and instead of comprising 12 / 26 = 46% of all D-type spherules, really comprise 0.8 / 14.8 ≈ 5%, in line with our predictions.

We see no difficulty in these microtektites being collected, despite their age. The majority of Australasian tektites show weak remanent magnetization and would be recoverable with a magnet (Pan *et al.* 2023). Given their age of 788 kyr, they would be buried in sediment, but perhaps by only 15 cm. The collection site lies 85 km from the nearest land (Manus Island), and sediment rates in deep ocean basins can be as low as ~10 cm / 800 kyr (Lyle, 2016). Notably, AATs recovered from the Indian Ocean (near 9°S, 77°E) were found in the topmost 15 cm of sediment (Prasad and Rao, 1990; Prasad *et al.* 1998, 1999; Prasad and Sudhakar, 1999). It has long been possible to extract cosmic spherules from the topmost tens of cm of the surface (Millard and Finkelman, 1970; Prasad *et al.* 2018), and the use of modern neodymium magnets only extends that range, especially if dragging the magnetic sled across the seafloor disturbed the topmost tens of centimeters of sediment layer.

In summary, as long as the thickness of sediment in the deep ocean site where Loeb *et al.* collected spherules is similar to the thickness (≈15 cm) of sediment in the Indian Ocean where other AATs were found (Prasad *et al.* 1999), we predict that Australasian microtektites should comprise about 22% of all spherules (micrometeorites and microtektites) at that site. They would have Mg/Si < 1/3 and be identified as Loeb et al. as "D-type" spherules. Loeb *et al*. claim these spherules "derive from the crustal rocks of a differentiated planet". The close match between the expected and observed frequency of such objects among all spherules allows us to identify that differentiated planet as Earth. We further predict that half of the "D-type" spherules would derive from lateritic sandstone layers and be resemble the iron-rich "BeLaU" subset of spherules. In the next section we carry out detailed geochemical comparisons to test this idea.

## GEOCHEMISTRY OF "BELAU" SPHERULES VS. MICROTEKTITES OF LATERITIC IRON

Having established that Australasian microtektites should be plentiful in the collection area at 1.3°S, 147.6°E, and should include a proportion of spherules from ferruginous or aluminous layers of lateritic sandstones, we discuss the geochemical evidence connecting these to the spherules. We begin with a review of laterites, then discuss the evidence from major and trace elements and other data that the "BeLaU" spherules derive from these layers.

**Laterites**

Laterites are highly weathered soils that see substantial vertical redistributions of elements over depths of a few meters to over 150 m. Different soil layers can be concentrated to ore grade in various elements, especially Al, Fe, Mn, Co, V, P, Ni, Cu and Au. The geochemistry of lateritic soils was reviewed by Nahon and Tardy (1992). Most commonly, the topmost layer, the 'iron crust', is pebbly and leached in alumina but enriched in iron. The alumina tends to be deposited, along with iron, in a layer below that. That layer is leached in silica and is Al- and Fe-rich. This 'cuirasse' layer is full of concretions of small spherules ('pisoliths') ranging in size from millimeters to centimeters, very similar to the hematite spherules, or 'blueberries', discovered by the Mars Exploration Rover on Mars (Squyres *et al.* 2004), as described by Jha *et al.* (2021). The material left in the near-surface pisolitic layers usually is dominated by Fe—as hematite, $Fe_2O_3$, or goethite, $FeO(OH)$—and Al—as gibbsite, $Al(OH)_3$, boehmite/diaspore, $AlO(OH)$, and kaolinite, $Al_2Si_2O_5(OH)_4$. Such layers may contain Ti as anatase ($TiO_2$) or ilmenite ($FeTiO_3$). Depending on which element dominates, these layers are sometimes called 'ferricretes' or Fe- or Al-duricrusts. These layers are of greatest relevance to the Papua New Guinea spherules. Under this is a layer of indurated conglomeratic iron crust, overlying a layer of soft nodular iron crust; these are leached in alumina and silica but enriched in iron. Descending several meters deeper, porosity decreases through a layer of clay-rich saprolite or 'saprock', with varying degrees of weathering that leach alkalis and silica from some of the parent rock fragments. At depths of meters to tens of meters is the parent rock.

As an example of the effects of lateritic weathering on sandstones, Jha *et al.* (2021) measured the abundances of major and trace elements in the pisolitic iron layer full of blueberry-like concretions (sizes < 1 mm to 25 mm), relative to the host Dhandraul sandstone, in Shankargarh, India. In this particular sandstone, comparing to the changes in relatively immobile $TiO_2$, $SiO_2$ was moderately depleted (by 22%) $Al_2O_3$ was highly depleted (by 91%), and MgO was completed depleted (100%), even as $Fe_2O_3$ was highly enriched, by +7500%. The Ca and Na were leached out (CaO depleted by 41%, $Na_2O$ depleted by 66%) but not K ($K_2O$ enriched by

1000%). MnO increased by 155% and $P_2O_5$ by 103%. These tremendous changes in composition occurred within tens of centimeters.

Trace elements were also concentrated in this layer, especially metals like Cr (+3550%), Zn (+2860%), Cu (+509%), and Ni (+490%), although not so much Co (-25%). Increases were generally seen in many volatile elements that would be lost if this sample were later melted: As (+3740%), Zn (+2860%), Sb (+2630%), Se (+1190%), Pb (+1170%), Ge (+1070%), Cd (+188%), Ga (+39%), Cs (+21%), Rb (-7%), Bi (-2%), Tl (-23%). Other notable enrichments include Be (+423%), V (+1270%), Ba (+458%), U (+184%) Sc (+275%), Nb (+18%), but slight depletions (relative to $TiO_2$) in Sr (-9%), Y (-18%), Th (-32%), Zr (-37%), Hf (-50%), and Li (-60%). Among the REEs, relative to $TiO_2$, the LREEs are enriched (by tens of percent) while the HREEs are depleted (by tens of percent).

These trends are merely representative of the chemical changes possible in a lateritic soil. Abundances will depend on the parent rock, the magnitude of the rainfall, whether the rainfall is continuous or undergoes a wet-dry seasonal cycle, the specific geochemistry of the area, etc. (Nahon and Tardy 1992). Within a particular laterite, abundances change continuously over lengthscales of tens of centimeters to meters.

**Major element patterns**

The enrichments of the uppermost layers of laterites in $Al_2O_3$ and Fe oxides, other metals, and the concomitant enrichments in Be, U, and LREEs like La, are of obvious relevance to the deep-sea "BeLaU" spherules collected near Papua New Guinea.

To make comparisons between the "BeLaU" spherules and lateritic sandstones, we must have the major element abundances of the spherules. Unfortunately, Loeb *et al.* did not report Si and O abundances measured by ICP-MS, so the $SiO_2$ content and oxidation state of the Fe in the spherules must be inferred. We do so for the case of S21, for which Loeb *et al.* supplied additional information from electron probe microanalysis (EPMA) spot analysis. We use the major elemental abundances reported by Loeb *et al.* for S21, including an apparent Si abundance of 0.65× CI, and infer it is mostly Fe and Fe oxides, and 14.8wt% $SiO_2$, 10.2wt% $Al_2O_3$, and only 0.37wt% MgO, assuming other species are these standard oxides. The extreme enrichment in Fe and Fe oxides relative to a UCC abundance of 5.0wt%, and the factor-of-4 depletion of $SiO_2$ and factor-of-7 depletion of MgO relative to standard UCC abundances (66.6wt%, 2.48wt%; Rudnick and Gao 2014) are immediately reminiscent of lateritic soils. Other oxide abundances are: CaO, 1.73wt%; $TiO_2$, 0.46wt%; $Na_2O$, 0.35wt%; $K_2O$, 0.16wt%; and $P_2O_5$, 0.23wt%. This, plus the abundances of trace elements, imply the spherule is roughly 71wt% Fe and Fe oxides ($FeO_T$), compared to 57.2wt% Fe. This implies oxygen in iron oxides is 25% the weight of the iron, roughly consistent with FeO or a mix of Fe and hematite. The EPMA spot analyses confirm some spots are pure Fe, while others seem more consistent with a mix of Fe and Fe oxides (Loeb *et al.* 2024). We assume that the $FeO_T$/Fe mass ratio is the same for the other "BeLaU" spherules, and we use this to calculate the $FeO_T$ and $SiO_2$ abundances.

In **Table 1** we report these calculated abundances for S21, the Al-rich spherule (25)IS13SPH5, and the average of the 12 "BeLaU" spherules. We also list the compositions for three lateritic soils in the Bolaven Plateau in southern Laos measured by Sanemastu *et al.* (2021). Sample 1607 is a sandstone laterite on the surface. Sample 1706A is a basaltic laterite measured at 0.1 m depth, and sample 1712C is a different basaltic laterite, measured at 2.0 m depth. Sanematsu *et al.* (2021) measured dozens of lateritic samples, and samples at the same location vary considerably (tens of wt%) with depth, and samples from nearby locations also even samples at similar depths at different locations can vary considerably (tens of wt%). These are representative only.

The "BeLaU" spherule S21 is broadly similar to sample 1607 from the surface layer of a lateritic sandstone, in that it is mostly Fe and Fe oxides, and in the range 10-17wt% $Al_2O_3$ and $SiO_2$, and highly depleted in MgO. S21 has more CaO, $Na_2O$ and $K_2O$ than the laterite, and less $TiO_2$ and $P_2O_5$. This suggests the Bolaven laterite is more $TiO_2$-rich and more alkali-depleted than the sandstone from which S21 derives, but otherwise the major element compositions are similar.

Likewise, the spherule (25)IS13SPH5, which is more Al-rich than many bauxite ores, resembles the basaltic laterite 1712C, in that both are 45-48wt% $Al_2O_3$, Fe and Fe oxides comprise the second-most abundant component (31wt% in the spherule, 19wt% in the soil), and $SiO_2$ and MgO are highly depleted. Again, the spherule is not as depleted in CaO, $Na_2O$ and $K_2O$ as the laterite.

Finally, the average of the "BeLaU" spherule compositions resembles the basaltic lateritic soil sample 1706A in their Fe +Fe oxide, $Al_2O_3$ and $SiO_2$ abundances. The average "BeLaU" spherule is depleted in MgO, but not by as much as 1706A. As with the other spherules, the average "BeLaU" spherule contains more CaO, $Na_2O$ and $K_2O$ than 1706A, about the same $P_2O_5$, and not as much $TiO_2$.

**Table 1:** Major oxide abundances of "BeLaU" spherules and lateritic soils.

|  | S21 | (25)IS13SPH5 | Avg. "BeLaU" | 1607 | 1712C | 1706A |
|---|---|---|---|---|---|---|
| Fe +Fe oxides | 71.4 | 31.3 | 49.4 | 54.6 | 19.1 | 42.5 |
| $Al_2O_3$ | 10.2 | 44.8 | 22.6 | 16.8 | 47.9 | 21.2 |
| $SiO_2$ | 14.8 | 4.7 | 18.2 | 10.4 | 1.67 | 15.7 |
| CaO | 1.7 | 13.0 | 4.3 | < 0.01 | < 0.01 | 0.01 |
| MgO | 0.37 | 1.47 | 1.16 | 0.04 | 0.13 | 0.06 |
| $TiO_2$ | 0.46 | 1.18 | 1.62 | 2.35 | 3.19 | 2.88 |
| $Na_2O$ | 0.35 | 1.78 | 1.33 | 0.02 | 0.03 | 0.02 |
| $K_2O$ | 0.16 | 1.45 | 0.85 | < 0.01 | 0.06 | 0.09 |
| $P_2O_5$ | 0.23 | 0.13 | 0.49 | 0.97 | 0.32 | 0.34 |
| MnO | 0.01 | 0.07 | 0.25 | 0.07 | 0.04 | 0.10 |

The major element patterns of the "BeLaU" spherules show considerable variations, but are overall similar to the abundances in the lateritic soils measured by Sanematsu *et al.* (2021) for the Bolaven Plateau sandstone and basalt laterites. We conclude that the "BeLaU" spherules derived from similar lateritic sandstones, with overall higher abundances of CaO and alkalis, and lower $TiO_2$ contents, suggesting a sandstone more plagioclase- and feldspar-rich and less basaltic in origin than the Bolaven Plateau soils.

**REE and trace element patterns**

A more stringent test of the hypothesis that the "BeLaU" spherules derived from lateritic sandstones is that their trace element abundances match, especially their rare earth element (REE) patterns. REE patterns are known to be fairly uniform across AATs, which is interpreted to mean that most of these tektites formed from well-mixed sediments (Ackerman *et al.* 2020). If the "BeLaU" spherules are also Australasian microtektites, they should have similar REE abundance patterns, meaning that abundances of La, Ce, etc., scaled to a normalizing REE element (e.g., Nd), should be identical between the spherules and AATs. The absolute concentrations should be broadly similar, but not necessarily identical, as lateritic soils can see substantial depletions in $SiO_2$, MgO, etc., reducing their masses by up to 80% and passively enriching REEs by factors of up to 5×.

Because the most compositional information is provided for spherule S21, we compare it to other AATs. The Australasian microtektites from the Larkman Nunavut moraine are well-characterized and relatively unweathered (van Ginneken *et al*. 2018). One in particular, LKN1153, is curiously $Al_2O_3$-rich (32.3wt%) and relatively $SiO_2$-poor (49.9wt%), somewhat suggestive of formation in an aluminous laterite despite its normal FeO abundance (2.3wt%). In **Figure 4**, we plot the abundances of both, normalized to the CI chondrite compositions of Anders and Grevesse (1989). We assume that the source rock of S21 experienced a greater degree of passive enrichment of REEs than the source rock of LKN1153. We multiply all the abundances of LKN1153 by an arbitrary factor of 1.4 to make a better comparison for REEs, although this would not necessarily be appropriate for major elements.

The agreement between S21 and LKN1153 is remarkable. Excluding major elements (Fe, Mg, Al, Si, Ti, Ca, Na, K, P, Mn), highly volatile elements (Pb, Rb, Cs) and metals (Fe, Co, Ni, Cr, Zn), the root-mean-square ratio between S21 and LKN1153 in 28 other elements (Li, Be, Sc, V, Cr, Sr, Y, Zr, Nb, Ba, REEs, Hf, Ta, Th, and U) is a factor of only 2.5. The REEs, in particular, match to within only tens of percent, and capture the trend of enrichment of LREEs over HREEs and the negative Eu anomaly. The abundances of Li and Be are exactly identical, and the abundances of the refractory lithophiles Sc, V, Sr, Y, Zr, Nb, Ba, Hf, Ta, Th, and U show similar enrichments. S21 is slightly more Sr- and U-rich than LKN1153, and LKN1153 more Zr, Nb, Hf-, Ta- and Th-rich, but the pattern is evident. The main difference to be noted is that S21 is more Fe-rich, and also enriched relative to LKN1153 in siderophile elements Co, Ni, Cr, and Zn. The volatile species Pb and Cs are clearly strongly depleted in both samples. S21 is more enriched in metal and more depleted in MgO and $SiO_2$ than LKN1153, but overall even the major element abundance pattern is similar between the two samples.

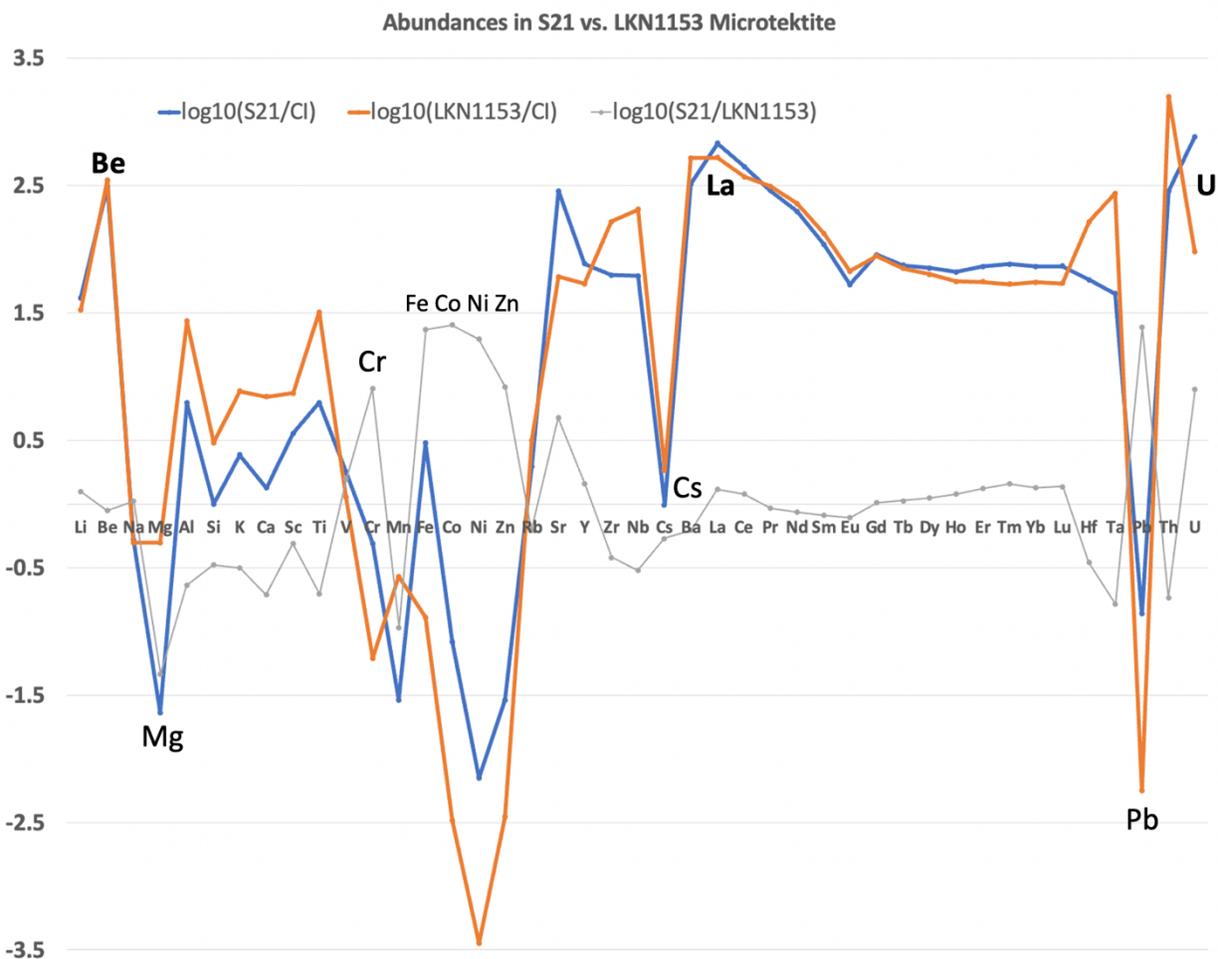

**Figure 4**. Spider diagram of major and trace elements (normalized to CI chondrites; Anders and Grevesse, 1989) in the "BeLaU" spherule S21 (blue) measured by Loeb *et al.* (2024), and the Australasian microtektite Larkman Nunavut (LKN) 1153 (orange), measured by van Ginneken *et al.* (2018). The LKN1153 abundances have been increased by a uniform factor of 1.4, to account for an assumed slightly higher passive enrichment in S21. S21 appears enriched in metals (Fe, Co, Ni, Cr, Zn) compared to LKN1153. CI-normalized abundances, and the ratios between the two samples (gray) are plotted on a logarithmic scale. Both S21 and LKN1153 are similarly enriched in Be, La, and U, depleted in Mg and Si, and depleted in the volatiles Pb, and Cs, and have nearly identical REE patterns. Spherule S21, and by extension the other "BeLaU" spherules, is simply a more metal-rich Australasian microtektite, consistent with formation in a lateritic sandstone.

**Other evidence**

Finally, we note other evidence linking the "BeLaU" spherules to the Australasian microtektites. Figures 4 and 5 of Loeb *et al.* (2024) show how different the morphologies are between the mostly-spherical cosmic spherules (especially those of I-type), and the irregularly shaped "D-type" spherules. Spherules like 19NMAG-52 and 19NMAG-50 appear to be compound, unlike all other cosmic spherules that form from ablation; if two completely molten droplets were to collide (which is improbable in the trail behind a meteor), they would likely coalesce into a single spherical droplet. Instead, these morphologies are more indicative of impact melts, a point emphasized by van Ginneken et al. (2018) about other spherules found in Antarctica. The morphology of "BeLaU" spherule S21 appears more consistent with a partially melted pisolitic texture (like a concretion of hematite 'blueberries'), and not a totally melted ablation spherule.

Finally, we note that the Fe isotopic compositions of the two measured "BeLaU" spherules, S10 and S21, are more consistent with microtektites than ablation spherules. Both fall exactly on the terrestrial fractionation line (TFL), as expected for materials from our Solar System, including Earth. They also lie remarkably close to the point $\delta^{56/54}Fe = 0‰$, $\delta^{57/54}Fe = 0‰$. S-type cosmic spherules formed by ablation of rocky meteoroids tend to have $\delta^{56/54}Fe = +1.4‰$ to +3.2‰, and similarly iron-rich I-type cosmic spherules tend of have $\delta^{56/54}Fe = +20‰$ to +36‰ (Engrand *et al.* 2005). The spherule S10 was measured by Loeb *et al.* (2023) to have $\delta^{56/54}Fe = -0.2‰$, and S21 to have $\delta^{56/54}Fe = +0.6‰$. This practically rules out origins as ablation spherules. Notably, though, the iron isotopic ratios of Australasian microtektites have been measured and found to lie on the TFL, mostly between $\delta^{56/54}Fe \approx -1‰$ and +2‰ (Chernonozhkin *et al.* 2021). The Fe isotopic ratios of the spherules S10 and S21, and by extension perhaps all "BeLaU" spherules, are completely consistent with them being Australasian microtektites.

## CONCLUSIONS

The site (1.3°S, 147.6°E) where Loeb *et al.* collected spherules is within the Australasian tektite strewn field, the most abundant source of near-surface tektites and microtektites in the world. The ejection of tektite from the impact site in the Gulf of Tonkin, in the direction of the Philippines, would have sent a high abundance of tektites to this site. In a column of sediment ≈15 cm thick deposited over the last 1 Myr, Australasian microtektites, all with Mg/Si < 1/3, would comprise about 22% of all spherules (microtektites plus micrometeorites) collected. This is exactly the proportion of spherules collected by Loeb *et al.* (2024) that they term "D-type".

The impactor is widely recognized to have impacted on a surface layer of laterite, and laterites are incorporated into Muong Nong layered tektites. Although not yet recognized as such, we estimate roughly ≈5% of Australasian microtektites should sample the Fe- and Al-rich uppermost layers of these laterites. Roughly 3% of all Australasian microtektites are known to be Al-rich; we suggest a similar number are Fe-rich but perhaps mistaken for I-type cosmic spherules. We note that Mizera *et al.* (2016) disfavored an impact site in southeast Asia specifically because it would be expected to produce Fe- and Al-rich microtektites; perhaps now the Fe-rich microtektites have been found. Because these iron-rich spherules would be more easily collected by magnets from deep-sea sediments compared to other microtektites (by about a factor of 15, in analogy to the proportions of I-type to S-type cosmic spherules), we predict roughly equal numbers of lateritic, Fe-rich spherules and otherwise typical microtektites. This matches the relative proportions of "BeLaU" and non-"BeLaU" "D-type" spherules collected by Loeb *et al.* (2024).

The major element compositions of the "BeLaU" spherules are entirely consistent with formation from the uppermost ferruginous and aluminous layers of lateritic sandstones. The compositions of laterites in southern Laos are broadly similar to the compositions of the "BeLaU" spherules, but which we infer derived from materials less depleted in Ca, Al, and Na, with less Ti, than the laterites in the Bolaven Plateau. This supports the idea that the impactor hit the Song Hong-Yinggehai basin in the Gulf of Tonkin.

Trace element patterns, including REE patterns, are expected to be uniform among Australasian tektites, presenting a test of the hypothesis. The trace element abundances are an excellent match to those of Australasian microtektites found in Antarctica. Both are highly, and identically, enriched in Be, La, and U (after normalizing to CI compositions and allowing for a factor of 1.4× passive enrichment).

The morphologies of the "D-type" spherules are compound and irregular and non-spherical. They are inconsistent with being ablation spherules, and are instead consistent with partially melted, pisolitic textures expected from the Fe-rich portions of lateritic soils.

The iron isotopic compositions of the spherules collected by Loeb *et al.* (2024) lie exactly (< 0.1‰) on the TFL, within 1‰ of the terrestrial value. This practically rules out an extrasolar origin, at the > 99.9% level. It also practically rules out an origin as ablation spherules from a meteoroid. Instead, the iron isotopic compositions are completely consistent with them being microtektites.

We conclude—based entirely on a comparison of existing literature with data Loeb *et al.* (2024) presented—that the D-type spherules are Australasian microtektites, and that the "BeLaU" subset derive from the Fe- and Al-rich layers of ferruginous sandstones. We predict that more example of such objects will be found on the seafloor wherever the sedimentation rates are < tens of cm over the last 1 Myr, anywhere along the line extending to the WNW, connecting through the Philippines to the impact site in southeast Asia.

These conclusions could have been reached by Loeb *et al.* (2024), but they did not collect the spherules with a plan for distinguishing them from other spherules in deep-sea sediments, nor consider ahead of time what sort of spherules they might encounter in this part of the world, being blind to the idea of tektites generally, and the Australasian tektites in particular. In their consideration of how rocky materials evolved, they never considered geochemical processes other than igneous fractionation. Rather than respond to the criticisms made (by Desch and Jackson, 2023) about their interpretations of Fe isotopic ratios (Loeb *et al.* 2023), Loeb *et al.*

(2024) ceased to discuss Fe isotopic ratios at all. Instead, they claim that the compositions of the "BeLaU" spherules are exotic, not seen anywhere else in the Solar System, and speculate about an unknown differentiated planet with a highly evolved crust, failing to recognize that that planet is Earth. This is not the first time that tektites have been attributed to the wrong planet: even after Spencer (1933) and Urey (1955) correctly identified them as terrestrial rock melted by impacts, Stair (1956) attributed them to a debris from a "Lost Planet" with a glassy surface. Barely justified in 1956, it is scientifically derelict in 2024 to fail even consider tektites and to publicly ascribe an extrasolar origin to a what is likely a piece of melted Earth.

*Declaration of Competing Interest*—The author declares that he has no known competing financial interests of personal relationships that could have appeared to influence the work reported in this paper.

*Acknowledgments*—The author thanks Ariel Anbar, Ben Fernando, Patricio Gallardo, Hilairy Hartnett, Rick Hervig, Alan Jackson, and Larry Nittler for useful conversations, and especially Laurence Garvie for sharing his expertise on tektites. The work herein benefitted from collaborations and/or information exchange within NASA's Nexus for Exoplanetary System Science research coordination network sponsored by NASA's Space Mission Directorate (grant 80NSSC23K1356, PI Steve Desch).